\documentclass[11pt,a4paper]{article}
\usepackage{jheppub}

\usepackage{graphicx}
\usepackage{setspace}

\begin{document}


\author[a]{Anton Alekseev,}



\author[a]{Olga Chekeres}



\author[b,c]{and Pavel Mnev}



\affiliation[a]{Department of Mathematics, University of Geneva,\\
 2-4 rue du Li\`evre, c.p. 64, 1211 Gen\`eve 4, Switzerland}
\affiliation[b]{Max Planck Institute for Mathematics,\\
Vivatsgasse 7, 53111 Bonn, Germany}
 \affiliation[c]{St. Petersburg Department of V. A. Steklov Institute of Mathematics of the Russian Academy of Sciences, \\
Fontanka 27, St. Petersburg, 191023 Russia}

\emailAdd{Anton.Alekseev@unige.ch}
\emailAdd{Olga.Chekeres@unige.ch}
\emailAdd{pmnev@pdmi.ras.ru}

\title{ Wilson surface observables from equivariant cohomology
}


\begin{abstract}   
{Wilson lines in gauge theories admit several path integral descriptions. The first one (due to Alekseev-Faddeev-Shatashvili) uses path integrals over coadjoint orbits. The second one (due to Diakonov-Petrov) replaces a 1-dimensional path integral with a 2-dimensional topological $\sigma$-model. We show that this $\sigma$-model is defined by the equivariant extension of the Kirillov symplectic form on the coadjoint orbit. This allows to define the corresponding observable on arbitrary 2-dimensional surfaces, including closed surfaces. We give a new path integral presentation of Wilson lines in terms of Poisson $\sigma$-models, and we test this presentation in the framework of the 2-dimensional Yang-Mills theory. On a closed surface,  our Wilson surface observable turns out to be nontrivial for $G$ non-simply connected (and trivial for $G$ simply connected), in particular we study in detail the cases $G=U(1)$ and $G=SO(3)$.
}

\keywords{Wilson line, Wilson surface, equivariant cohomology, 2d Yang-Mills, gauge theories}

\dedicated{Dedicated to Ludwig Faddeev \\
on the occasion of  his $3^4$th anniversary}


\end{abstract}

\maketitle


\section{Introduction}

Wilson line observables play an important role in gauge theories. Such an observable is defined by a closed curve $\Gamma$ and a representation of the gauge group $R$. It is described by the formula
$$
W^R_\Gamma = {\rm Tr}_R \, P \, \exp\left( \int_\Gamma A\right),
$$
where $A$ is  a gauge field. 

Wilson lines admit several interesting path integral presentations. The first one is due to Alekseev-Faddeev-Shatashvili \cite{AFS}, and it involves the auxiliary field $b=g\lambda g^{-1}$, where $g$ is a group element defined on the curve $\Gamma$ and $\lambda$ is the label of the coadjoint orbit which corresponds to the representation $R$. The formula reads
$$
W^R_\Gamma(A)= \int \mathcal{D} g \, e^{i S_\lambda(A,g)},
$$
where the auxiliary action $S_\lambda(A,g)$ is given by the expression
$$
S_\lambda(A,g)=\int_\Gamma\, {\rm Tr} \, \lambda(g^{-1}dg + g^{-1} Ag) = \int_\Gamma {\rm Tr} \, b(dgg^{-1} +A).
$$

The second presentation is due to Diakonov-Petrov, and a path integral of an auxiliary 2-dimensional field theory is defined on the
surface $\Sigma$ bounding the curve $\Gamma$,
$$
W^R_\Gamma(A)= \int \mathcal{D} g \, e^{i \int_\Sigma {\rm DP}_\lambda(A,g)} .
$$
In \cite{DP}, this presentation was used in the discussion of the area law for Wilson lines in a gauge theory with confinement.

In this article, we study the relation between the two path integral presentations. Our first result is a beautiful formula for the 
Diakonov-Petrov Lagrangian
\begin{equation} \label{key}
{\rm DP}_\lambda(A,g) = {\rm Tr} \, b(F_A - (d_Agg^{-1})^2),
\end{equation}
where $F_A$ is the field strength and $d_Ag=dg+Ag$. With this definition, the Diakonov-Petrov action can be defined on closed surfaces
and on surfaces with multiple boundary components.

Our second result is the interpretation of the formula \eqref{key} in terms of equivariant cohomology. 
Coadjoint orbits carry the canonical Kirillov symplectic form. We show that the expression ${\rm DP}_\lambda(A,g)$
is the equivariant extension of this symplectic form corresponding to the action of the gauge group on the orbit.

Next, we use the Diakonov-Petrov formula to give a new path integral presentation of Wilson lines in terms of the Poisson $\sigma$-model.
In our case, the Poisson $\sigma$-model reduces to a 2-dimensional BF-theory
$$
S = \int {\rm Tr} \, b(d(A+\alpha) + (A+\alpha)^2),
$$
where the field $b$ takes values in the coadjoint orbit, $A$ is the external gauge field and $\alpha$ is the auxiliary gauge field of the Poisson
$\sigma$-model. In case when $A$ is a connection on a non trivial bundle, the expression $A+\alpha$ defines another connection on the same bundle.

We test our theory of Wilson surfaces in the case of the 2-dimensional Yang-Mills (2YM) theory. In this case, the YM theory is exactly solvable, and one can see
the effects of adding a Wilson surface observable in the partition function.  We look in detail at the cases of $G=U(1)$, $G=SU(2)$ and $G=SO(3)$. For $G=SU(2)$, a simply connected group, a Wilson surface of any spin doesn't change the partition function for a closed surface. That is, the corresponding observable turns out to be trivial. For $G=U(1)$, a Wilson surface observable carries a parameter $\lambda \in \mathbb{R}$. For $\lambda \in \mathbb{Z}$ the observable is trivial, and it is nontrivial for $\lambda \notin \mathbb{Z}$.  In the case of $G=SO(3)$, the Wilson surface observable is a $\mathbb{Z}_2$-valued topological invariant of the $SO(3)$-bundle, and
and we observe an interesting relation
$$
e^{i\int DP_{1/2} (A,g)} = \sqrt{e^{i\int DP_1 (A,g)}}
 $$
between surface observables of integer and half-integer spin. Observables of integer spin are trivial (as in the case of $G=SU(2)$). But the observables of half-integer spin prove to be nontrivial, and one can see how they change the 2YM partition function on a closed surface.

Recently, $\mathbb{Z}_2$-valued topological invariants appeared in the theory of topological insulators \cite{Kane, Ryu, Gaw, Gaw2, Kapusta}.
A possible relation of  our results with these invariants  is to be explored. 

Gauge invariant quantities assigned to surfaces rather than curves were extensively studied in literature \cite{Ganor, Chen, Chepel, Cat}. Our approach differs from most other approaches in that we have no higher gauge fields in the game. Our starting point is the standard gauge field which is a 1-form with values in a Lie algebra. The surface observable is obtained by using a non-abelian Stokes formula from a Wilson line observable. It is surprising that this construction allows an invariant formulation suitable for surfaces with many boundary components and for closed surfaces. It is also surprising that even for closed surfaces one may construct nontrivial observables (as in the case of $G=U(1)$ or $G=SO(3)$).

\vskip 0.2cm

{\bf Acknowledgements.} We are grateful to Y. Barmaz, P. Severa, S. Shatashvili and C. Schweigert for useful discussions.
The research of A.A. and O.C. was supported in part by the grant MODFLAT of the European Research Council (ERC), the grants
152812 and 141329  and the NCCR SwissMAP of the Swiss National Science Foundation. P. M. acknowledges partial support of RFBR Grant No. 13-01-12405-ofi-m. 
We would like to thank the referee of this article for very useful remarks and questions.

\section{A simple example: the case of $G=U(1)$}\label{2}

		We start with a simple example of the first Chern class of a principal circle bundle over an orientable surface $\Sigma$. Let $P \to \Sigma$ be a principal $U(1)$-bundle, and let $\mathfrak{\tilde{a}}\in \Omega^1(P)$ be a connection on $P$. Then, $F_\mathfrak{\tilde{a}} = d\mathfrak{\tilde{a}}$ is the curvature of $P$ and the first Chern form. It is basic and descends to a 2-form on $\Sigma$. If $\Sigma$ is closed, then
\begin{equation}
c_1(P) = \frac{1}{2\pi} \, \int_\Sigma F_\mathfrak{\tilde{a}} 
\end{equation}
is an integer called the first Chern number of $P$. One can view the defining equation for the curvature, $F_\mathfrak{\tilde{a}} = d \mathfrak{\tilde{a}}$ as the definition of the 1-dimensional Chern-Simons form \cite{Freed1, Freed2},
$$
F_\mathfrak{\tilde{a}}= d \,  {\rm CS}_1(\mathfrak{\tilde{a}}),
$$
where ${\rm CS}_1(\mathfrak{\tilde{a}})=\mathfrak{\tilde{a}}$.

Assume that the surface $\Sigma$ is connected, orientable and has a nontrivial boundary $\Gamma= \partial \Sigma \neq \emptyset$. Then, the circle bundle $P$ is necessarily trivial. Let's choose a global section $\sigma: \Sigma \to P$ and define the gauge field $a = \sigma^* \mathfrak{\tilde{a}}$ so that $F_a = \sigma^* F_\mathfrak{\tilde{a}}$.\footnote{To distinguish between various connections appearing in the paper, in 1-dimensional CS-theory we use the Gothic letter $\mathfrak{\tilde{a}}$ for the connection 1-form on the principal bundle and the latin letter $\it{a}$ for the 1-form representing this connection on the manifold.} Then, one can define a  quantity $S(a)$ associated to $\Sigma$ via
$$
S(a)= \int_\Sigma F_a.
$$
We can think of this expression as of the simplest surface observable associated to the surface $\Sigma$. Using the Stokes formula, we obtain
\begin{equation}  \label{paradox}
S(a)= \int_\Sigma  F_a = \int_\Sigma da = \int_{\Gamma} a.
\end{equation}
Let $\phi: \Sigma \to U(1)$ and consider the gauge transformation $a^\phi=a+d\phi$. Then, the curvature $F_a$ is gauge invariant, and so is the expression for the surface observable $\int_\Sigma F_a$.  

However, in the expression $\int_\Gamma a$ the gauge invariance is lost! Indeed,
\begin{equation}\label{Chern}
\int_\Gamma a^\phi = \int_\Gamma (a + d\phi) = \int_\Gamma a + \left( \phi(2\pi) - \phi(0) \right).
\end{equation}
In general, $\phi(2\pi) - \phi(0) = 2\pi n$ with $n \in \mathbb{Z}$. The fact that the left hand side of \eqref{paradox} is gauge invariant and the right hand side is not may appear as a contradiction. But there is a solution to this puzzle: for gauge transformations defined on the surface $\Sigma$ (that is, $\phi: \Sigma \to U(1)$) we have $\phi(2\pi) = \phi(0)$, and the extra term $\phi(2\pi) - \phi(0)$ always vanishes.

If we use the expression $S(a)=\int_\Gamma a$ and admit arbitrary gauge transformations defined on $\Gamma$, it is the exponential $\exp(iS(a))$ which becomes gauge invariant, while $S(a)$ is not gauge invariant in general.

\section{Two path integral presentations of Wilson lines}

In this Section, we shall discuss an example of the bulk-boundary correspondence based on the path integral description of Wilson lines in gauge theory. 

\subsection{The Alekseev-Faddev-Shatashvili (AFS) formula}

Let $G$ be a compact connected Lie group, $\mathfrak{g}={\rm Lie}(G)$ its Lie algebra, and  ${\rm Tr}$ an invariant scalar product on $\mathfrak{g}$. Consider a manifold $N$, and let $P$ be a principal $G$-bundle $P \to N$.  A connection $\mathcal{A}$\footnote{For general non-abelian case we use the capital Gothic letter $\mathcal{A}$ to denote the connection 1-form on the principal bundle and the capital letter $A$ for the 1-form representing this connection on the manifold.} is given by a $\mathfrak{g}$-valued 1-form on $P$.

Let $\Gamma \subset N$ be a closed curve.
The restriction of $P$ to a closed curve $\Gamma$ is always trivial (since the group $G$ is connected). Hence, we can use a connection 1-form $A\in \Omega^1(\Gamma, \mathfrak{g})$ instead of $\mathcal{A}$.

A Wilson line \cite{Wilson, Giles, Witten3} is an observable defined by the holonomy of the gauge field $A$ along a closed curve $\Gamma$, embedded in $N$, and a finite dimensional representation $R$ of $G$. It is given by the formula
\begin{equation}
W_\Gamma^R(A)= {\rm Tr}_R \, P\exp\left( \int_\Gamma A^R\right),
\end{equation}
where $A^R$ is a matrix-valued 1-form obtained by taking $A$ in the representation $R$,
\begin{equation}
A^R = A^a_i \, t_a^R \, dx^i.\footnote{Summation is implied over the repeated indices.}
\end{equation}
The path ordered exponential $P \exp$ stands for  the holonomy of the gauge field,
\begin{equation}
P\exp\left( \int_\Gamma A^R\right) = 1 + \int_\Gamma A^R + \frac{1}{2!} \int_{s_1 > s_2} A^R(s_1) \wedge A^R(s_2) + \dots
\end{equation}
where $s_1, s_2$ are parameters on the curve $\Gamma$.

Irreducible finite dimensional representations of $G$ are in one-to-one correspondence with integral coadjoint orbits $O \subset \mathfrak{g}^*$ \cite{Kirusha}. In more detail, an irreducible representation is uniquely determined by its highest weight $\lambda \in \mathfrak{h}^*$, where $\mathfrak{h} \subset \mathfrak{g}$ is a Cartan subalgebra of $\mathfrak{g}$ \cite{Zhelob, Bott}. Then, one can associate to $\lambda$ the orbit of the coadjoint action in the space $\mathfrak{g}^*$. By abuse of notation, we write it as a matrix conjugation
\begin{equation}
{\rm Ad}^*_g (\lambda) = g\lambda g^{-1},
\end{equation}
and denote the coadjoint orbit by
\begin{equation}
O_\lambda=\{ g\lambda g^{-1}; g \in G\}.
\end{equation}

A path integral presentation of Wilson lines described by Alekseev-Faddeev-Shatashvili (see also \cite{Elitzur, Bala, Nielsen, CB, DP2}) works as follows. Let $b: \Gamma \to \mathfrak{g}^*$ be an auxiliary field defined on the curve $\Gamma$ and taking values in $O_\lambda$. In addition, we introduce a field $g: \Gamma \to G$ which takes values in $G$, with the property $b(s)=g(s) \lambda g(s)^{-1}$. Then,
\begin{equation}
W_\Gamma^R = \int {\mathcal{D}} g \, e^{i S_\lambda(A,g)},
\end{equation}
where 
\begin{equation}
S_\lambda(A, g) =  \int_\Gamma \, {\rm Tr} \, \lambda(g^{-1} dg + g^{-1} A g) =
\int_\Gamma \, {\rm Tr} \, b(dg g^{-1} + A).
\end{equation}
We can introduce the differential form
\begin{equation}\label{AFS action}
{\rm AFS}_\lambda(A,g) = {\rm Tr} \, \lambda(g^{-1} dg + g^{-1} A g) =  {\rm Tr} \, b(dg g^{-1} + A),
\end{equation}
such that
$$
S_\lambda(A,g) = \int_\Gamma {\rm AFS}_\lambda(A,g).
$$

This action is invariant with respect to the following gauge transformations with parameter $h: \Gamma \to G$:
$$ g \mapsto hg, \,\, A \mapsto A^h=hAh^{-1} - dhh^{-1}, \,\, b \mapsto hbh^{-1}.$$

Indeed, it is easy to check that  
\begin{spacing}{1.45}
$
\begin{array}{lll}
AFS_\lambda (A^h, hg) & = &  {\rm Tr} \, \lambda((hg)^{-1} d(hg) + (hg)^{-1} (hAh^{-1} - dhh^{-1}) hg) \\
& = & {\rm Tr} \, \lambda(g^{-1} dg + g^{-1} A g)\\
& = & AFS_\lambda(A,g).
\end{array}
$
\end{spacing}

Hence,  $S_\lambda (A^h, hg) = S_\lambda(A,g)$.

It is interesting to consider another class of gauge transformations
$$ g \mapsto gt^{-1}, \, t \in H_\lambda,  
$$
where $H_\lambda$ is the subgroup of $G$ preserving $\lambda$ under the coadjoint action:
$$
H_\lambda = \{ h \in G; \,\, h \lambda h^{-1} =\lambda \} .
$$
For $\lambda \in \mathfrak{h}^*$ generic, $H_\lambda$ is the Cartan subgroup of $G$.
However, $S_\lambda(A,g)$  is not invariant under these transformations. 
Instead, it acquires an additional term:

\begin{equation} 
S_\lambda(gt^{-1}, A)  =  S_\lambda(g, A) - \int_{\Gamma} {\rm Tr}\lambda(dtt^{-1}) =  S_\lambda(g, A)  - 2 \pi {\rm Tr}(\lambda \vec{n}),
\end{equation}
where $\vec{n}=\int_\Gamma dtt^{-1}$. The components of $\vec{n}$ are the winding numbers of the map $t: \Gamma \to H_\lambda$.
The exponential $e^{ikS_\lambda(g, A)}$ is gauge invariant if 
$$
k\, {\rm Tr} \, (\lambda \vec{n}) \in \mathbb{Z} 
$$
for vectors $\vec{n}\in \mathbb{Z}^r$.
If $\lambda$ is an integral weight, then $Tr(\lambda \vec{n})$ is always an integer. This requires the coefficients $k$ to be quantized.

\subsection{The Diakonov-Petrov (DP) formula}

The second presentation of the Wilson line is due to Diakonov and Petrov who came up with the following formula:
\begin{equation}
{\rm DP} = d \, {\rm AFS}_\lambda(A,g).
\end{equation}
Expanding the form $d \, {\rm AFS}_\lambda(A,g)$ we obtain the expression for the Diakonov-Petrov Lagrangian in terms of  a matrix-valued 1-form $A$ and a matrix-valued function $g$:

\begin{spacing}{1.45}
$
\begin{array}{lll} 
d \, {\rm AFS}_\lambda(A,g) & = & d\,\, {\rm Tr} \, \lambda(g^{-1} dg + g^{-1} A g) \\
& = & {\rm Tr} \, \lambda(d g^{-1} dg + d g^{-1} Ag + g^{-1} dA g - g^{-1} A dg) \\ 
& = & {\rm Tr} \, \lambda (-g^{-1} dg g^{-1} dg - g^{-1} dg g^{-1} Ag + g^{-1} dA g - g^{-1} Ag g^{-1} dg) \\
& = & {\rm Tr} \, \lambda (g^{-1} F_A g - \frac{1}{2} g^{-1}[A,A] g - \frac{1}{2} [g^{-1} dg , g^{-1} dg] - [g^{-1} dg , g^{-1} Ag]) \\
& = & {\rm Tr} \, b ( F_A - \frac{1}{2} [dg g^{-1} + A , dg g^{-1} + A] ) \\
& = & {\rm Tr} \, b \left( F_A - (d_A g g^{-1})^2 \right) \\
& = & {\rm DP}_\lambda(A,g),
\end{array}
$
\end{spacing}
where we used $dA = F_A - \frac{1}{2} [A,A]$, $dg^{-1} = -g^{-1} dg g^{-1}$ and $d_A g = dg + Ag$.

Now consider a surface $\Sigma$ bounded by the curve $\Gamma$. One can use the Stokes formula to give a new expression for the 
action $S_\lambda(A,g)$,
$$
S_\lambda(A,g)= \int_\Gamma {\rm AFS}_\lambda(A,g) = \int_\Sigma {\rm DP}_\lambda(A,g) = 
\int_\Sigma {\rm Tr} \, b \left( F_A - (d_A g g^{-1})^2 \right).
$$
The right hand side is manifestly gauge invariant whereas the expression $\int_\Gamma {\rm AFS}_\lambda(A,g)$ is not. The explanation is similar to the one that we encounter in the case of $G=U(1)$: for gauge transformations $g: \Sigma \to G$, both expressions for the action are gauge invariant; which is not necessarily the case for the integral of ${\rm AFS}_\lambda(A,g)$ if one considers a gauge transformation $g: \Gamma \to G$. 

The path integral is taken over all configurations of the field $g$ including all boundary values.

\section{Equivariant cohomology approach}

The main result of this Section is the equivariant cohomology \cite{At, Berline, Super, Meinrenken2} interpretation of the Diakonov-Petrov formula.


\subsection{The Weil model of equivariant cohomology}
 Let $M$ be a smooth  manifold, $G$ be a compact connected Lie group acting on $M$ and $\mathfrak{g}$ be the Lie algebra of $G$. 
To every $\xi \in \mathfrak{g}$ we associate the fundamental vector field $\xi_M \in \mathfrak{X}(M)$.

The de Rham differential $d$, contractions $\imath_\xi^M$ and Lie derivatives $L_\xi^M$ act on the space of differential forms $\Omega^*(M)$, and satisfy the relations
$$
[\imath_\xi^M, \imath_\eta^M] =0, \quad [L_\xi^M, \imath_\eta^M] = \imath_{[\xi, \eta]}^M, \quad [L_\xi^M, L_\eta^M]=L_{[\xi, \eta]}^M,
\quad [d, \imath_{\xi}^M] =L_{\xi}^M, \quad [d, L_{\xi}^M]=0, \quad [d,d]=0,
$$
where $[-,-]$ stands for a supercommutator.

Together, they form a Lie superalgebra  $\mathcal{G}$.  One natural class of representations of $\mathcal{G}$ are spaces of differential forms $\Omega^*(M)$. Another representation is the Weil algebra:
\begin{equation}
W_G:= S{\mathfrak{g}^*} \otimes \wedge {\mathfrak{g}^*},
\end{equation}
which is constructed by taking the product of the symmetric and exterior algebras of the dual space to $\mathfrak{g}$.
The Weil algebra is graded  by assigning degree 2 to generators of $S\mathfrak{g}^*$ and degree 1 to generators of $\wedge \mathfrak{g}^*$,
\begin{equation}
W^l_G = \oplus_{j+2k=l} S^k {\mathfrak{g}^*} \otimes \wedge^j \mathfrak{g}^*.
\end{equation}
It is convenient to introduce elements $a, f \in W_G \otimes \mathfrak{g}$ constructed as follows: $a$ is a element in $\wedge^1 \mathfrak{g}^* \otimes \mathfrak{g}$ defined by the canonical pairing between $\mathfrak{g}$ and $\mathfrak{g}^*$. Similarly, $f \in S^1\mathfrak{g}^* \otimes \mathfrak{g}$.

The superalgebra $\mathcal{G}$ acts on $W_G$ as follows:
\begin{align}
 \label{dadf}
			da &= f - \frac{1}{2}[a, a], \hskip 1.0cm   df = [f, a].\\
                \imath_\xi^W a &= \xi, \hskip 2.3cm \imath_\xi^W f = 0.\\
			L_{\xi}^W f &= [f, \xi], \hskip 1.6cm   L_{\xi}^W a = [a, \xi].
			\end{align}

One can think of $a$ as a universal connection on a principal $G$-bundle. Then,  the first formula in \eqref{dadf} gives the standard definition of the curvature and the second one is the Bianchi identity. 

As representations of $\mathcal{G}$ are carried by both $\Omega^*(M)$ and $W_G$ the diagonal action on the tensor product can be defined.
Thus $d$, $\imath_\xi$, $L_\xi$ act on $\Omega^*(M) \otimes W_G$ as follows:

\begin{spacing}{1.45}
\begin{equation}
\begin{array}{lll} 

L_{\xi} &=& L_{\xi}^M \otimes 1 + 1 \otimes L_{\xi}^W, \\
\imath_{\xi} &=& \imath_{\xi}^M \otimes 1 + 1 \otimes \imath_{\xi}^W, \\
d &=& d \otimes 1 + 1 \otimes d.
\end{array}
\end{equation}
\end{spacing}

The space $\Omega^*_G (M)$ of equivariant forms on $M$  is then defined as the basic part of $\Omega^*(M) \otimes W_G$:

$$ \Omega^*_G (M) :=\{ \alpha \in \Omega^*(M) \otimes W_G | L_{\xi} \alpha = 0, \imath_\xi \alpha = 0 \}. $$

And the Weil model of equivariant cohomology on $M$ is defined as:
\begin{equation}
H^*_G(M)  = H^*( \Omega_G ^* (M), d\otimes 1 + 1\otimes d).
\end{equation}

\subsection{The form ${\rm DP}_\lambda(A,g)$ as an equivariant cocycle}
Now we consider in more detail the Diakonov-Petrov action, where the form

\begin{equation}\label{DPform}
{\rm DP}_\lambda(A,g)= {\rm Tr} \, \lambda (g^{-1} F_A g - \frac{1}{2} g^{-1}[A,A] g - \frac{1}{2} [g^{-1} dg , g^{-1} dg] - [g^{-1} dg , g^{-1} Ag])
\end{equation}
is of particular interest. Here $A$ is the gauge field on  $N$ and $F_A$ is its curvature.  We can now construct an equivariant differential form on the orbit ${\rm O}_\lambda$ by replacing $A$ with $a$ and $F_A$ with $f$. The resulting element has the form
\begin{equation}
{\rm DP}_\lambda(a,g) = {\rm Tr} \, \lambda (g^{-1} f g - \frac{1}{2} g^{-1}[a,a] g - \frac{1}{2} [g^{-1} dg , g^{-1} dg] - [g^{-1} dg , g^{-1} ag])
\end{equation}
 It is a well-known fact that the coadjoint orbits $O_\lambda \subset \mathfrak{g}^*$ (discussed in the previous Section) are symplectic manifolds $(O_\lambda, \varpi_O)$. Here the symplectic form $\varpi_O$ on $O_\lambda$ can be identified as one of the terms in ${\rm DP}_\lambda(a,g)$: 

\begin{equation}\label{KirForm}
\varpi_O = - {\rm Tr} \, \lambda(g^{-1}dg)^2= - {\rm Tr} \, b(dg g^{-1})^2.
\end{equation}
This is a closed and non-degerate 2-form also known as the Kirillov form. 

Our first claim is that ${\rm DP}_\lambda(a,g)$ is equivariantly closed. Indeed,
$$
{\rm DP}_\lambda(a,g) = d \,{\rm AFS}_\lambda(a,g)=d \, {\rm Tr} \, \lambda(g^{-1} dg + g^{-1} a g).
$$
Hence,  $d \,{\rm DP}_\lambda(a,g)= 0$.

Futhermore, applying the combined contraction gives:
\begin{equation}
\imath_{\xi} {\rm DP}_\lambda(a,g) = {\rm Tr} \, b (  - \frac{1}{2} [-\xi g g^{-1} + \xi , dg g^{-1} + a] + \frac{1}{2} [dg g^{-1} + a , - \xi g g^{-1} + \xi] ) = 0,
\end{equation}
where we have used that $\imath_{\xi}(dg) = -\xi g$.

Closedness and horizontality of ${\rm DP}_\lambda(a,g)$ imply vanishing of its Lie derivative:
\begin{equation}
L_{\xi} {\rm DP}_\lambda(a,g) = (\imath_{\xi} d + d \imath_{\xi}) {\rm DP}_\lambda(a,g) = 0.
\end{equation}

 The two conditions $L_{\xi} {\rm DP}_\lambda(a,g) = 0, \, \imath_{\xi} {\rm DP}_\lambda(a,g) = 0$ being satisfied,  ${\rm DP}_\lambda(a,g)$ is an equivariant differential form on the coadjoint orbit $O_\lambda$.  Since it is equivariantly closed, we can view it as an equivariant extension of the Kirillov symplectic form.

\section{Poisson $\sigma$-model formula for a Wilson line}

In this Section, we introduce another description for a Wilson line observable in terms of a 2-dimensional path integral.

Recall that for a Poisson manifold $(M, \pi)$ with Poisson structure
$$
\pi = \frac{1}{2} \, \pi^{ij} \, \frac{\partial}{\partial x^i} \wedge \frac{\partial}{\partial x^j},
$$
the corresponding Poisson $\sigma$-model \cite{Ikeda, Str, Schaller} is defined by the action
$$
S^\pi (X, \alpha) = \int_\Sigma \left( \alpha_i dX^i + \frac{1}{2} \pi^{ij}(X) \alpha_i \wedge \alpha_j\right).
$$
Here $X: \Sigma \to M$ is a map of the surface $\Sigma$ to the target space $M$, $X^i = x^i \circ X$ are its components,
and $\alpha_i$ are 1-forms on $\Sigma$ representing gauge fields of the Poisson $\sigma$-model. In case when $\pi^{ij}$ is
invertible, one can integrate out the fields $\alpha$ to obtain an integral of the symplectic form $\omega_{ij}=(\pi^{-1})^{ji}$:
\begin{equation}\label{Omega}
S_\omega = \frac{1}{2} \int_\Sigma \omega_{ij} dX^i \wedge dX^j .
\end{equation}

In particular, when $\Sigma$ is a surface with boundary the path integral is taken over all $X$, including boundary values, and over the auxiliary fields $\alpha$ which vanish on the boundary:
 
$$
\int_{\alpha|_{\partial \Sigma}=0} DX D\alpha \, e^{iS^\pi(X,\alpha)} = \int DX e^{iS_\omega(X)}.
$$

Now let the target space be $\mathcal{O}_\lambda$, a coadjoint orbit of $G$ passing through the point $\lambda$. (For this part of the discussion we avoid unnecessary complexity and work with a trivial $G$-bundle over the world-sheet $\Sigma$.) The form $\omega$ is given by \eqref{KirForm}, and the action \eqref{Omega} reads
$$
S_{\varpi_O} = - \int_\Sigma {\rm Tr} \, \lambda (g^{-1}dg)^2 = - \int_\Sigma {\rm Tr} \, b (dgg^{-1})^2.
$$ 
To construct the Poisson $\sigma$-model on the coadjoint orbit, we have to add the auxiliary gauge fields $\alpha \in \Omega^1(\Sigma, \mathfrak{g})$. The action $S^\pi$ is given by
\begin{equation}\label{BF}
S^\pi (b, \alpha) =\int_\Sigma {\rm Tr} \, b(d\alpha + \alpha^2).
\end{equation}
And again, for $\Sigma$ a surface with boundary we impose the condition that the auxiliary field $\alpha$ vanishes on the boundary.\footnote{If further on we want to perform gluing on the surface observable, instead of fixed $\alpha|_{\partial\Sigma}=0$ we then allow a family of boundary conditions $\alpha|_{\partial\Sigma}=\alpha_0$ and integrate over $\alpha_0$ for gluing.} 
In fact, in the expression \eqref{BF} one can identify the 2-dimensional BF-theory \cite{Birm} with an extra condition that $b$ belongs to a fixed coadjoint orbit. 

To make the relationship between $S_{\varpi_O}$ and $S^\pi$ more transparent, we rewrite the latter as follows:
\begin{equation}\label{intout}
 \int_\Sigma {\rm Tr} \, b(d\alpha + \alpha^2) = \int_\Sigma {\rm Tr} \, b\left( (dgg^{-1} + \alpha)^2 - (dgg^{-1})^2\right),
 \end{equation}
where we have used integration by parts and the fact that ${\rm Tr} b [dgg^{-1}, \alpha] = {\rm Tr} [b, dgg^{-1}] \alpha = - {\rm Tr} (db) \alpha$. It is now clear that integrating out $\alpha$ yields the expression for $S_{\varpi_O}$.

Recall that the Diakonov-Petrov form \eqref{DPform} is  an equivariant extention of the Kirillov form and thus the Diakonov-Petrov action 
$$
S_{DP}=\int_\Sigma {\rm Tr} b(F_A - (d_Agg^{-1})^2)
$$
can be viewed as a version of the action $S_{\varpi_O}$ interacting with the external gauge field $A$. Introducing an auxiliary gauge field $\alpha$ and performing the transformations similar to \eqref{intout} we define the Poisson $\sigma$-model version of this action by
\begin{spacing}{1.45}
\begin{equation}\label{PSM}
\begin{array}{lll}
S_\sigma (b, A, \alpha) & = & \int_\Sigma {\rm Tr} \, b \left(F_A + (d_Agg^{-1} + \alpha)^2 - (d_Agg^{-1})^2\right) \\
& = & \int_\Sigma {\rm Tr} \, b \left( F_A + d_A\alpha + \alpha^2\right) \\
& = & \int_\Sigma {\rm Tr} \, b\left( d(A+\alpha) + (A+\alpha)^2\right).
\end{array}
\end{equation}
\end{spacing}
From the first line, it is obvious that integrating out $\alpha$ yields the Diakonov-Petrov action. Surprizingly, the final expression coincides
with the Poisson $\sigma$-model for the coadjoint orbit \eqref{BF}, but with the new gauge field $A+\alpha$ combining the background field $A$ and the gauge field of the Poisson $\sigma$-model $\alpha$.

Up to now all the argumentation above works in the case when $A$ is a gauge field on a trivial $G$-bundle over $\Sigma$. If this is not the case, a further analysis is required.
In fact, for a nontrivial geometric setup all the formulas still hold. However, one should be more precise about where the fields take values. Let $P\to \Sigma$ be a (possibly nontrivial) principal
$G$-bundle and $\mathcal{A} \in \Omega^1(P, \mathfrak{g})$ be a connection on $P$. The curvature $F_{\mathcal{A}}=d\mathcal{A}+\mathcal{A}^2$ belongs to $\Omega_{\rm hor}^2(P, \mathfrak{g})^G$,
the space of horizontal $G$-invariant 2-forms with values in $\mathfrak{g}$. Since the expression
$$
\int_\Sigma {\rm Tr} \, b F_{\mathcal{A}}
$$
makes part of the action $S_\sigma (b, \mathcal{A}, \alpha)$, the field $b$ must take values in $\Omega^0_{\rm hor}(P, \mathfrak{g})^G$, that is in $G$-invariant functions on $P$ with values in $\mathfrak{g}$.
The expression ${\rm Tr} \, b F_{\mathcal{A}}$ is then a horizontal invariant 2-form on $P$, and it descends to $\Sigma$. In a similar fashion  $\alpha \in \Omega_{\rm hor}^1(P, \mathfrak{g})^G$, and  interestingly the combination $\mathcal{A}+\alpha$ defines a new connection on $P$.

\section{The interpretation of surface observables in terms of topology of principal bundles}\label{5.1}

In this Section, we give an interpretation of the surface observables in terms of the first Chern class of the bundle \cite{Cheeger-Simons, CS, WittenDij, Milnor} with the structure group $H_\lambda$. 

Let $N$ be a manifold (space-time), $P\to N$ be a principle $G$-bundle over $N$, 
$\Sigma \subset N$ be a submanifold of $N$,  and $P|_\Sigma \to \Sigma$ be the restriction of the principal bundle $P$ to $\Sigma$. Assume that over $\Sigma$ the structure group $G$ of $P$ reduces to a subgroup $H \subset G$. That is, $\Sigma$ carries a principal $H$-bundle $Q \to \Sigma$, and there is an $H$-equivariant inclusion $ \it{i}: Q \to P|_\Sigma$. 

The bundle $Q \to \Sigma$ is a pull-back of the universal bundle $EH\to BH$ under the map $\sigma: \Sigma \to BH$.
It induces a map in cohomology $\sigma^*: H^*(BH) \to H^*(\Sigma)$. Since $\Sigma$ is 2-dimensional, we are particularly interested in the cohomology group $H^2(BH) \cong {\rm char}(\mathfrak{h}) \subset \mathfrak{h}^*$. Here ${\rm char}(\mathfrak{h})$ is the set of characters of the Lie algebra $\mathfrak{h}$,

\begin{equation} \label{[]}
{\rm char}(\mathfrak{h})=\{ \lambda \in \mathfrak{h}^*; \,\,  \langle \lambda, [x,y] \rangle =0 \,\,  {\rm for \, all} \, x,y \in \mathfrak{h} \} .
\end{equation}

For a closed surface $\Sigma$, we will show that the surface observable of the previous sections is given by
%
%
$$
\int_\Sigma \sigma^* c_\lambda,
$$
where $c_\lambda \in H^2(BH)$ is the image of  $\lambda \in {\rm char}(\mathfrak{h})$.

In more detail, the form ${\rm DP}_\lambda(a,g)$ defined on a universal G-bundle can be viewed as an element of $\Omega_G^2(P \times Q , \mathfrak{g})$. Let $\pi: \mathfrak{g} \to \mathfrak{h}$ be an $H$-equivariant projection from $\mathfrak{g}$ to $\mathfrak{h}$, and let
\begin{equation}
\mathfrak{a} = \pi(g^{-1} dg + g^{-1} \mathcal{A} g).
\end{equation}
This is an element of $\Omega^1(P \times Q, \mathfrak{h})$. We will show that
\begin{equation}
{\rm DP}_\lambda(\mathcal{A},g) = {\rm Tr}\, \lambda (d\mathfrak{a} + \mathfrak{a}^2).
\end{equation}

First, observe that the condition \eqref{[]} implies ${\rm Tr} \, \lambda (\mathfrak{a}^2) = 0$. Indeed,  $\mathfrak{a}^2 = \frac{1}{2} [\mathfrak{a},\mathfrak{a}]$ and $\mathfrak{a}$ takes values in $\mathfrak{h}$. Thus, we are interested in the expression 
$${\rm Tr} \, \lambda \, d\mathfrak{a} = {\rm Tr} \, \lambda\, (d\mathfrak{a} + \mathfrak{a}^2).$$

Second, recall the structure of the invariant pairing between the elements of $\mathfrak{h}$ and its dual $\mathfrak{h}^*$, $\lambda \in \mathfrak{h}^*$ being an element of the dual to the Cartan subalgebra of $\mathfrak{g}$. One can view $\mathfrak{g}$ as a direct sum of the subalgebra $\mathfrak{h}$ and its invariant complement $\mathfrak{p}$ (that is,
$[\mathfrak{h}, \mathfrak{p}] \subset \mathfrak{p}$):
$$\mathfrak{g} \cong \mathfrak{h} \oplus \mathfrak{p}.$$

Then the invariant product between two elements is defined in the following way:
$$ {\rm Tr} \, (\lambda x) = \langle \lambda , x \rangle \, \, for \, \, \lambda \in \mathfrak{h}^*, \, x \in  \mathfrak{h},$$
$$ {\rm Tr} \, (\lambda y) = 0 \, \, for \, \, \lambda \in \mathfrak{h}^*, \, y \in  \mathfrak{p}.$$
And the product of $\lambda$ with an element of $\mathfrak{g}$ under projection to $\mathfrak{h}$ is the same as the product of $\lambda$ with this element itself:
$$ {\rm Tr} \,( \lambda \, \pi (x+y)) = {\rm Tr} \, (\lambda x)= {\rm Tr} \, (\lambda (x+y) )= \langle \lambda , x \rangle.$$

Thus the following direct computation proves our claim:
\begin{spacing}{1.45}
$\begin{array}{lll}
{\rm Tr} \, \lambda (d\mathfrak{a} + \mathfrak{a}^2) & = & {\rm Tr}\, \lambda d\mathfrak{a} \\
& =&  {\rm Tr} \,\lambda \pi(d(g^{-1} dg + g^{-1} \mathcal{A} g)) \\
& = & {\rm Tr}\, \lambda \pi (g^{-1} F_{\mathcal{A}} g - \frac{1}{2} g^{-1}[\mathcal{A},\mathcal{A}] g - \frac{1}{2} [g^{-1} dg , g^{-1} dg] - [g^{-1} dg , g^{-1} \mathcal{A}g]) \\
& = & {\rm Tr}\, \lambda \pi (g^{-1} F_{\mathcal{A}} g - (g^{-1} d_{\mathcal{A}} g)^2) \\
& =  & {\rm Tr} \,\lambda (g^{-1} F_{\mathcal{A}} g - (g^{-1} d_{\mathcal{A}} g)^2) \\
& =&  {\rm DP}_\lambda(\mathcal{A},g).
\end{array}$
\end{spacing} 
Hence, we conclude
\begin{equation}\label{Class}
\int_\Sigma {\rm DP}_\lambda(\mathcal{A}, g) = \int_\Sigma {\rm Tr} \,\lambda (d\mathfrak{a} + \mathfrak{a}^2) = \int_\Sigma \sigma^* c_\lambda,
\end{equation}
as required.
Thus, our more sophisticated definition for a surface observable in the case of $G$ non-abelian is structurally the same as the one in the case of $G=U(1)$. And the value of the observable can be identified with the first Chern number $c_1(Q)$ of the bundle $Q \subset P$ over $\Sigma$.


\section{Application to 2-dimensional Yang-Mills theory}

In this Section, we study the effects of adding a Wilson surface in the 2D Yang-Mills theory. Since this theory is exactly solvable \cite{Migdal, Bralic, Kazak, Kazak2, Kazak3, Blau2, Fine, Fine2, Rus, Gross}, one obtains explicit formulas for partition function in the presence of a Wilson surface.

\subsection{2-dimensional Yang-Mills theory}


Recall that the action of the Yang-Mills theory is given by


\begin{equation}
 S_{YM}(A)  = \frac{1}{4 e^2} \int_{\Sigma} {\rm Tr} F_A \ast F_A,
 \end{equation}
where $*$ is the Hodge dual defined by the metric on the surface $\Sigma$. In the first order formalism, this action can be rewritten as
\begin{equation}
S(A,B)= \int_{\Sigma} {\rm Tr} ( BF_A + \frac{e^2}{2} B^2 d^2\sigma),
\end{equation}
where $B$ is an auxiliary field and $d^2\sigma$ is the area element on $\Sigma$.

The canonical quantization of the theory on a cylinder $S^1 \times \mathbb{R}$ gives rise to the Hilbert space with a basis $\chi_R(w)$, where
$R$ runs through the set of irreducible representations of $G$, and 
$$
w = {\rm P exp} \int_{S^1} A
$$
is the holonomy of $A$ around $S^1$. The partition function of the theory on an orientable  surface of genus $g$ with $r$ boundary components reads
\begin{equation}
Z(\tau, w_1, \dots, w_r)=\sum_R d_R^{2-2g-r} e^{- \tau C_2(R)} \chi_R(w_1) \dots \chi_R(w_r).
\end{equation}

Here $\tau = \frac{e^2}{2}\sigma$, $\sigma$ the area of the surface,  $w_1, \dots, w_r$ are holonomies of $A$ around the boundary components, the sum runs over all irreducible representations of $G$, and $C_2(R)$ is the quadratic Casimir in the representation $R$. In particular, for a closed surface one obtains
\begin{equation}
Z(\tau) = \sum_R d_R^{2-2g} e^{- \tau C_2(R)}.
\end{equation}
For more details see e. g. \cite{Witten5, Moore}.

\subsection{U(1)} 
As a warm up example, we consider the case $G=U(1)$.  The partition function for 2D-YM with Wilson surface can be obtained through Hamiltonian formalism. The idea is to construct the new Hamiltonian which already contains the Wilson surface. The formula for the partition function would be:
 \begin{equation} 
Z_{\lambda}(\tau) = {\rm Tr} \, e^{-itH_{\lambda}},
\end{equation}
where
$H_{\lambda}$ 
is the Hamiltonian of the theory perturbed by a Wilson surface operator: 

Let $w = e^{i\int_0^L dx A_1}$ be a holonomy around a curve of a constant time slice. The representations of $U(1)$ are labelled by integers $n \in \mathbb{Z}$. The characters of the representations are $\chi_n(w) = w^n$. 
Then the eigenvalues of the Hamiltonian are given by:
$$
H_{\lambda} \chi_n(w) =  \frac{e^2}{2} L  (n-\lambda) ^2 \chi_n(w),
$$
where $L$ is the length of the cylinder on which the quantization takes place.
And the partition function becomes:
\begin{equation}\label{Z_U(1)}
Z_{\lambda}(\tau) = \sum_{n \in \mathbb{Z}} e^{ -i\tau (n - \lambda )^2}.
\end{equation}
Notice that the label of the representation n gets a shift by $-\lambda$ due to the presence of the Wilson surface.	

On the other hand, recall that in the case of $G=U(1)$ the Wilson surface observable $S(A)$ coincides with the first Chern class of the 
corresponding $U(1)$-bundle. Hence, after adding a factor $\exp(i S(A))$ in the definition of the partition function the calculation of the path integral yields
\begin{equation}\label{U_Chern}
Z_\lambda(\tau) = \int DA \, e^{iS_{YM} + iS_{\lambda}}  = \beta \, \sum_{m \in \mathbb{Z}} e^{ i\pi^2m^2/\tau + i 2\pi \lambda m}.
\end{equation}	
where $\beta$ is a constant and the meaning of the 
parameter $m$ is the first Chern number of the $U(1)$-bundle over $\Sigma$.
	
  A nice observation can be made that the partition function computed from the Hamiltonian formalism (the sum over the representations) is related through the Poisson resummation to the partition function obtained from the functional integral formalism (the sum over the first Chern number of the U(1)-bundles over $\Sigma$): 
$$
\sum_{n \in \mathbb{Z}} e^{ -i\tau (n - \lambda )^2} = \sqrt{\frac{\pi}{i\tau}} \, \sum_{m \in \mathbb{Z}} e^{ i\pi^2m^2/\tau + i 2\pi \lambda m}.
$$
From this resummation we can deduce the value of the constant $\beta = \sqrt{\frac{\pi}{i\tau}}$.

\subsection{$G$ arbitrary}

We shall use the Poisson $\sigma$-model action representing the Wilson surface observable. In the first order formalism, the total action reads
\begin{equation}
S=\int_\Sigma {\rm Tr}  \left( BF_A + \frac{e^2}{2} B^2 d^2\sigma + b F_{A+\alpha}\right),
\end{equation}
where $F_{A+\alpha}=d(A+\alpha) + (A+\alpha)^2$ is the field strength of the gauge field $(A+\alpha)$ calculated in equation \eqref{PSM}. Hence, we obtain two non-interacting BF-theories. The first one has a Hamiltonian $\frac{e^2}{2} {\rm Tr} B^2$, and the field $B$ can vary in $\mathfrak{g}^*$. The second one has vanishing Hamiltonian, and the field $b$ takes values in the coadjoint orbit passing through $\lambda$. The Hilbert space of such a system has a basis
\begin{equation}
\psi_R(A, \alpha)= \chi_R(w(A)) \chi_{R_\lambda}(w(A+\alpha)),
\end{equation}
where $R$ runs through irreducible representations of $G$, $R_\lambda$ is the irreducible representation corresponding to the coadjoint orbit passing through $\lambda$, and
$$
w(A) = {\rm P exp} \int_{S^1} A \, ,  \hskip 0.6cm w(A+\alpha) = {\rm P exp} \int_{S^1} (A+\alpha).
$$
Since only the first BF-theory contributes in the Hamiltonian, its eigenvalue on $\psi_R(A, \alpha)$ is given by $C_2(R)$, as in the pure YM theory.

On a surface of genus $g$ with $r$ boundary components, the partition function reads
\begin{equation}
Z_\lambda(\tau, A, \alpha))=
d_{R_\lambda}^{-r} \sum_R d_R^{2-2g-r} e^{- \tau C_2(R)} \chi_R(w_1(A)) \dots \chi_R(w_r(A)) \chi_{R_\lambda} (w_1(A+\alpha)) \dots \chi_{R_\lambda}(w_r(A+\alpha))
\end{equation}
In particular, for a closed surface we obtain exactly the same expression as for the pure YM theory:
$$
Z_\lambda(\tau)=\sum_R d_R^{2-2g} e^{- \tau C_2(R)} =Z(\tau).
$$

The Wilson surface observable which has no effect on the partition function is hardly of any interest.
However, while considering the cases when $R_\lambda$ is a projective representation of the gauge group $G$, we obtain nontrivial results. 
The difference is transparent in the $U(1)$ example discussed in the previous subsection. If in formulas \eqref{U_Chern} and \eqref{Z_U(1)} $\lambda$ is an integral weight of the $U(1)$ representations, the shift of the Chern number in the exponential is not visible. If we allow $\lambda$ to be a weight of a representation  of $\mathbb{R}$ (that is, a projective representation of $U(1)$) the partition function changes.
The same principle is applies to a more interesting example of the gauge group $SO(3)$ and its universal cover $SU(2)$ which we discuss in more detail in the next section.

\subsection{SU(2) and SO(3) in Hamiltonian formalism}

We now apply the general formalism of the previous section to the case of $G=SU(2)$ and $G=SO(3)$. We keep the notation $Z(\tau)$ for the partition function of the free 2YM theory and $Z_\lambda (\tau)$ for the theory with a Wilson surface of spin $\lambda$.

For \textbf{G=SU(2)}, irreducible representations are labeled by integer and half-integer spins $j$. The dimension of such a  representation is $2j+1$, and the quadratic Casimir element is $\left( j +\frac{1}{2}\right)^2$. The corresponding formula for the partition function on the surface of genus $g$ with $r$ boundary components reads
\begin{equation}
\begin{array}{lll}
Z_\lambda(\tau, A, \alpha) &=&
(2\lambda +1)^{-r} \sum_{j\in\mathbb{Z}^{\geq 0}/2} (2j+1)^{2-2g-r} e^{- \tau (j+1/2)^2 } \times \\
& & \times \chi_j(w_1(A))...\chi_j(w_r(A)) \chi_{\lambda} (w_1(A+\alpha))...\chi_{\lambda}(w_r(A+\alpha)).
\end{array}
\end{equation}
Here the sum is over non negative integer and half-integer values of $j$, and the characters of irreducible representations are defined by formula
\begin{equation}
\chi_j\left(
\begin{array}{cc}
c & 0 \\
0 & c^{-1}
\end{array}
\right)
= \frac{c^{j+1/2} - c^{-(j+1/2)}}{c^{1/2} - c^{-1/2}} .
\end{equation}
For the closed surface, we get the following result 
\begin{equation}
Z_{\lambda, SU(2)}(\tau)=
 \sum_{j\in \mathbb{Z}^{\ge0}/2} (2j+1)^{2-2g} e^{- \tau (j+1/2)^2 } 
\end{equation}
which coincides with the partition function $Z(\tau)$ without Wilson surface. 

In the case of \textbf{G=SO(3)}, the partition function for a closed surface may be calculated as a sum of contributions of the trivial and nontrivial bundles over $\Sigma$ \cite{Witten2}:
\begin{equation}
Z_{{\rm SO}(3)}(\tau)  =  Z^{\rm triv}(\tau) + Z^{\rm nontriv}(\tau).
\end{equation}
Here the contribution of the trivial bundle 
$$
Z^{\rm triv}(\tau) = \frac{1}{2} \, Z_{SU(2)}(\tau).
$$
To obtain the contribution of the nontrivial representation, we consider the partition function of the surface with a disc removed, in such a way that one boundary component appears:
\begin{equation}
Z(\tau, w) = \sum_{j\in \mathbb{Z}^{\ge0}/2} (2j+1)^{1-2g} e^{-\tau (j+1/2)^2} \chi_j(w),
\end{equation}
and then put $w=-e$ (the nontrivial central element of $SU(2)$) to get 
\begin{spacing}{1.45}
\begin{equation}
\begin{array}{lll}
Z^{\rm nontriv}(\tau) &=& \frac{1}{2} Z(\tau, -e)\\
&=& \frac{1}{2} \sum_{j\in \mathbb{Z}^{\ge0}/2} (2j+1)^{1-2g} e^{-\tau (j+1/2)^2} \chi_j(-e)\\
& =& \frac{1}{2}  \sum_{j\in \mathbb{Z}^{\ge0}/2} (-1)^{2j} (2j+1)^{2-2g} e^{-\tau (j+1/2)^2}.
\end{array}
\end{equation}
\end{spacing}

The resulting partition function for $SO(3)$ yields
$$
Z_{{\rm SO}(3)}(\tau) =  \sum_{j\in \mathbb{Z}^{\ge0}} (2j+1)^{2-2g} e^{- \tau (j+1/2)^2 },
$$
where the sum is now over non negative integer spins, as expected.

Now we apply the same procedure to a partition function with Wilson surface of spin $\lambda$. For the trivial part we obtain
\begin{equation}
Z_\lambda^{\rm triv}(\tau)= \frac{1}{2} \, \sum_{j\in \mathbb{Z}^{\ge0}/2} (2j+1)^{2-2g} e^{- \tau (j+1/2)^2 },
\end{equation}
as before. And the nontrivial contribution yields:
\begin{spacing}{1.45}
\begin{equation}
\begin{array}{lll}
Z_\lambda^{\rm nontriv}(\tau) & = & \frac{1}{2}\, Z_\lambda(\tau, -e,-e) \\
&=& \frac{1}{2(2\lambda +1)} \sum_{j\in \mathbb{Z}^{\ge0}/2} (2j+1)^{1-2g} e^{-\tau (j+1/2)^2} \chi_j(-e) \chi_\lambda(-e) \\
&=& \frac{(-1)^{2\lambda}}{2}  \sum_{j\in\mathbb{Z}^{\ge0}/2} (-1)^{2j} (2j+1)^{2-2g} e^{-\tau (j+1/2)^2}.
\end{array}
\end{equation}
\end{spacing}
If the spin of the Wilson surface observable $\lambda$ is  an integer, summing up the two contributions reproduces the same answer as for the theory without Wilson surface. 
However, if $\lambda$ is half-integer, we have $(-1)^{2\lambda}=-1$ and partition function is changed by the presence of the observable:
\begin{equation}
Z_{\lambda, {\rm SO}(3) }(\tau) = \sum_{j\in \frac{1}{2}+\mathbb{Z}^{\ge0}} (2j+1)^{2-2g} e^{- \tau (j+1/2)^2 },
\end{equation}
where the sum is over $j$ which now take only half-integer values!

Thus the formula for the partition function in 2YM with a Wilson surface observable for the gauge group $G=SO(3)$ is given by

\begin{spacing}{1.45}
\begin{equation}\label{SO3}
\begin{array}{lll}
&Z_{\lambda, {\rm SO}(3) }(\tau)&= \sum_{j\in \mathbb{Z}^{\ge0}} (2j+1)^{2-2g} e^{- \tau (j+1/2)^2 }, \hskip 1.1cm \lambda \in \mathbb{Z}^{\ge0},\\

&Z_{\lambda, {\rm SO}(3) }(\tau) &= \sum_{j\in \frac{1}{2}+\mathbb{Z}^{\ge0}} (2j+1)^{2-2g} e^{- \tau (j+1/2)^2 },  \hskip 0.8cm \lambda \in \frac{1}{2}+\mathbb{Z}^{\ge0}.
\end{array}
\end{equation}
\end{spacing}

Note, that we observe a shift of the representation labels $j$, similar to the $U(1)$ case formula \eqref{Z_U(1)}.

\subsection{Topological approach to SU(2) and SO(3) partition functions}

In this Section, we would like to confirm by path integral computations the results obtained previously for $G=SO(3)$ . Recall that for pure $SO(3)$ Yang-Mills theory 
on a closed surface we have
$$
Z(\tau) =  \int \mathcal{D}A \mathcal{D}g \, e^{iS_{YM}} =  Z^{\rm triv}(\tau) + Z^{\rm nontriv}(\tau),
$$
where on the right hand side we split the contributions of trivial and nontrivial $SO(3)$-bundles over the surface.

For the theory with Wilson surface observable, the partition function is
\begin{equation}
Z_{\lambda,SO(3)}(\tau) = \int \mathcal{D}A \mathcal{D}g \, e^{iS_{YM} \, + i\int DP_\lambda (A,g)}.
\end{equation}
It turns out that the Wilson surface factor $e^{i\int DP_\lambda (A,g)}$ only depends on the topological type of the bundle and not on the particular gauge field choices on this bundle. 

Recall the topological description of the observable given in Section \ref{5.1} and assume that the field $g$ defines a circle subbundle $Q \subset P$. Then according to the formula \eqref{Class}, the term of the action corresponding to the Wilson surface is
 \begin{equation}
 S_\lambda =  \int DP_\lambda (A,g) =  \int \, {\rm Tr} \, \lambda d\mathfrak{a} =  2\pi \lambda c_1(Q).
\end{equation}
Here $\mathfrak{a} \in \Omega^1(Q, {\rm Lie}(S^1))$ is a connection 1-form on the subbundle $Q$ taking values in the Lie algebra of $S^1$, $c_1(Q) = \frac{1}{2\pi}\int_\Sigma d\mathfrak{a}$ is the first Chern number of the $S^1$-subbundle over $\Sigma$ and $\lambda$ is the coefficient of the Wilson surface with the meaning of spin.

Note that for 2YM the $SO(3)$-bundle and the $S^1$-subbundle are over the same base-space $\Sigma$. Hence the $SO(3)$-bundle $P\to \Sigma$ is completely determined by its subbundle $Q\to \Sigma$. In more detail, since $S^1\subset SO(3)$ we are allowed to use the transition functions $\phi \in S^1$ of $Q$ as transition functions of $P$. The $SO(3)$-bundle is then given by $P=SO(3)\times_{S^1} Q$.\footnote{This is a particular instance of  the construction of relative bundles (see \cite{Steen}) used to describe topological defects in \cite{Schweigert}.} 
 
 The transition function of $Q$  is $\phi: S^1 \to S^1$, and the transition function of $P$ is $\hat{\phi}: S^1 \to S^1 \to SO(3)$. It is obtained from $\phi$ by composition with the embedding of the maximal torus $S^1 \to SO(3)$.  The equivalence class of the bundle $P$  is determined by the homotopy class $[\hat{\phi}] \in \pi^1(SO(3))=
 \mathbb{Z}_2 = \{+1, -1\}$. 
 At the same time the equivalence class of the subbundle $Q$ is determined by $[\phi] \in \pi^1(S^1)=
 \mathbb{Z}$ which corresponds to an integer winding number and hence to the first Chern number $c_1(Q)$. The relation between the winding number and $c_1(Q)$ is transparent from the formula \eqref{Chern}. The induced map between the bundles equivalence classes $ \mathbb{Z} \to \mathbb{Z}_2$ maps even Chern numbers to the trivial element of $\mathbb{Z}_2$ and odd Chern numbers to the nontrivial one \cite{Hatcher}. Then an $SO(3)$-bundle defined by an $S^1$-subbundle with an even Chern class is trivial, and an $SO(3)$-bundle defined by a subbundle with an odd Chern class is necessarily nontrivial. 
 
 Another interpretation of the map $ \mathbb{Z} \to \mathbb{Z}_2$ between the bundles equivalence classes can be given. For $\Sigma$ connected and orientable, an $SO(3)$-bundle $P\to \Sigma$ is completely determined, up to a bundle isomorphism, by its 2nd Stiefel-Whitney class $w_2\in H^2(\Sigma,\mathbb{Z}_2)\simeq \mathbb{Z}_2$, while a $U(1)$-bundle $Q$ is determined by its 1st Chern class $c_1\in H^2(\Sigma,\mathbb{Z})\simeq \mathbb{Z}$. For $P=Q\times_{U(1)}SO(3)$, these classes are related by $w_2(P)=c_1(Q)\mod 2$.
 
 The Wilson surface factor in the functional integral then gives:
 
 \begin{spacing}{1.45}
\begin{equation}
\begin{array}{lll}
&e^{i2\pi \lambda c_1(Q)}& = +1, \hskip 1.1cm {\rm P \, \, is \, \, trivial}, \\

&e^{i2\pi \lambda c_1(Q)}& = +1, \hskip 1.1cm {\rm P \, \, is \, \, nontrivial}, \, \, \lambda \in \mathbb{Z}^{\ge0}, \\
&e^{i2\pi \lambda c_1(Q)}& = -1,  \hskip 1.1cm {\rm P \, \, is \, \, nontrivial}, \, \lambda \in \frac{1}{2}+\mathbb{Z}^{\ge0}.
\end{array}
\end{equation}
\end{spacing}
Note that the exponentials for surface observables for $\lambda=1/2$ and for $\lambda=1$ are related as follows:
$$
e^{i\int DP_1 (A,g)}=\left( e^{i\int DP_{1/2} (A,g)}\right)^2
$$
or 
$$
e^{i\int DP_{1/2} (A,g)} = \sqrt{e^{i\int DP_1 (A,g)}}.
 $$
That is, the surface observable for half-integer spin is a nontrivial square root of the observable for integer spin.

For $\lambda$ half-integer the nontrivial part of the partition function acquires  a factor $-1$ and becomes
$$
Z_\lambda^{\rm nontriv}(\tau) = - \frac{1}{2} \, \sum_{j\in\mathbb{Z}^{\ge0}/2} (-1)^{2j} (2j+1)^{1-2g} e^{-\tau (j+1/2)^2}.
$$

 This leads to the following formula:

\begin{spacing}{1.45}
\begin{equation}
\begin{array}{lll}
 &Z_{\lambda,SO(3)}(\tau) &=  Z^{\rm triv}(\tau) + Z^{\rm nontriv}(\tau), \hskip 0.8cm \lambda \in \mathbb{Z}^{\ge0}\\
&Z_{\lambda,SO(3)}(\tau) &=  Z^{\rm triv}(\tau) - Z^{\rm nontriv}(\tau), \hskip 0.8cm \lambda \in \frac{1}{2}+\mathbb{Z}^{\ge0}.
 \end{array}
\end{equation}
\end{spacing}
Summing up the trivial and nontrivial contributions reproduces exactly the result of the equation \eqref{SO3}. 
For the $SU(2)$ gauge group the bundle $P$ is necessarily trivial and the Wilson surface does not affect the 2YM partition function.



\begin{thebibliography}{99}


\bibitem{AFS} A. Alekseev, L. Faddeev, S. Shatashvili,
\textit{Quantization of symplectic orbits of compact Lie groups by means of the functional integral, 
J. Geom. Phys} \textbf{5} (1988) 391.

\bibitem{DP} D. Diakonov, V. Petrov,
\textit{Non-Abelian Stokes theorem and quark-monopole interaction} [hep-th/9606104],
Published version: \textit{Nonperturbative approaches to QCD, Proceedings of the Internat. workshop at ECT*}, Trento, July 10-29, 1995, D.Diakonov (ed.), PNPI (1995).

\bibitem{Kane} C. L. Kane, E. J. Mele,
\textit{$Z_2$ topological order and the quantum spin Hall effect,
Phys. Rev. Lett.} \textbf{95} (2005) 146802 [cond-mat/0506581].

\bibitem{Ryu} S. Ryu, C. Mudry, H. Obuse, A. Furusaki,
\textit{$Z_2$ Topological Term, the Global Anomaly, and the Two-Dimensional Symplectic Symmetry Class of Anderson Localization,
Phys. Rev. Lett.} \textbf{99} (2007) 116601 [cond-mat/0702529].

\bibitem{Gaw} D. Carpentier, P. Delplace, M. Fruchart, K. Gawedzki, C. Tauber,
\textit{Construction and properties of a topological index for periodically driven time-reversal invariant 2D crystals,
Nucl. Phys.} \textbf{B 896} (2015) 779 [arXiv:1503.04157].

\bibitem{Gaw2} D. Carpentier, P. Delplace, M. Fruchart, K. Gawedzki,
\textit{Topological index for periodically driven time-reversal invariant 2D systems,
Phys. Rev. Lett.} \textbf{114} (2015) 106806 [arXiv:1407.7747].

\bibitem{Kapusta} A. Kapustin,
\textit{Bosonic topologial insulators and paramagnets: a view from cobordisms},  arXiv:1404.6659.

\bibitem{Ganor} O. Ganor,
\textit{Six-dimensional tensionless strings in the large N limit,
Nucl. Phys.} \textbf{B 489} (1997) 95 [hep-th/9605201].


\bibitem{Chen} B. Chen, W. He, J.-B. Wu and L. Zhang,
\textit{M5-branes and Wilson surfaces,
JHEP} \textbf{08} (2007) 067 [arXiv:0707.3978].

\bibitem{Chepel} I. Chepelev,
\textit{Non-Abelian Wilson Surfaces, 
JHEP} \textbf{02} (2002) 013 [hep-th/0111018].

\bibitem{Cat} A.Cattaneo, C. Rossi,
\textit{Wilson surfaces and higher dimensional knot invariants},
\textit{Commun.Math.Phys.} \textbf{256} (2005) 513 [math-ph/0210037].

\bibitem{Freed1} D. S. Freed, 
\textit{Classical Chern-Simons theory, Part 1,  Adv.Math.} \textbf{113} (1995) 237 [hep-th/9206021].

\bibitem{Freed2} D. S. Freed,
\textit{Classical Chern-Simons theory, Part 2, Houston J. Math.} \textbf{28} (2002) 293.


\bibitem{Wilson} K. Wilson,
\textit{Confinement of quarks,
Phys. Rev.} \textbf{D 10} (1974) 2445.

\bibitem{Giles} R. Giles,
\textit{Reconstruction of gauge potentials from Wilson loops,
Phys. Rev.} \textbf{D 24} (1981) 2160.


\bibitem{Witten3} E. Witten,
\textit{Quantum Field Theory and the Jones Polynomial,
Comm. Math. Phys.} \textbf{121} (1989) 351. 

\bibitem{Kirusha} A. A. Kirillov, 
\textit{Lectures on the orbit method}, Graduate Studies in Mathematics \textbf{64}, Providence, RI: American Mathematical Society (2004).

\bibitem{Zhelob} D. P. Zhelobenko,
\textit{Compact Lie Groups And Their Representations}, 
Translations of Mathematical Monographs \textbf{40}, American Mathematical Society (1978).

\bibitem{Bott} R. Bott,
\textit{The Geometry and Representation Theory of Compact Lie Groups}, 
in \textit{Representation Theory of Lie Groups}, London Mathematical Society Lecture Note Series, Cambridge University Press \textbf{34} (1979).

\bibitem{Bala} A. P. Balachandran, S. Borchardt, A. Stern,
\textit{Lagrangian And Hamiltonian Descriptions of Yang-Mills Particles,
Phys. Rev.} \textbf{D 17} (1978) 3247.

\bibitem{Nielsen} H. B. Nielsen, D. Rohrlich,
\textit{A Path integral to quantize Spin,
Nuci. Phys.} \textbf{B 299} (1988) 471.

\bibitem{DP2} D. Diakonov, V. Petrov, 
\textit{Phys. Lett.} \textbf{B 224} (1989) 131.

\bibitem{Elitzur} S. Elitzur, G. Moore, A. Schwimmer, N. Seiberg,
\textit{Remarks on the Canonical Quantization of the Chern-Simons-Witten Theory,
Nucl. Phys.} \textbf{B 326} (1995) 108.

\bibitem{CB} C. Beasley,
\textit{Localization for Wilson Loops in Chern-Simons Theory},
in J. Andersen, H. Boden, A. Hahn, and B. Himpel (eds.) \textit{Chern-Simons Gauge Theory: 20 Years After}
\textit{AMS/IP Studies in Adv. Math.} \textbf{50} (2011), \textit{Adv. Theor. Math. Phys.} \textbf{17} (2013) 1 [arXiv:0911.2687].


\bibitem{At} M.F. Atiyah and R. Bott,
\textit{The moment map and equivariant cohomology,
Topology} \textbf{23} (1984) 1.

\bibitem{Super} V. W. Guillemin, S. Sternberg, 
\textit{Supersymmetry and Equivariant de Rham Theory}, 
Springer Berlin Heidelberg (1991).

\bibitem{Berline} N. Berline, E. Getzler, M. Vergne, 
\textit{Heat Kernels and Dirac Operators},
Grundlehren Text Editions, Springer Berlin Heidelberg (2004). 


\bibitem{Meinrenken2} E.Meinrenken,
\textit{Equivariant cohomology and the Cartan model},
http://www.math.toronto.edu/mein/research/enc.pdf 

\bibitem{Ikeda} N. Ikeda,
\textit{Two-dimensional gravity and nonlinear gauge theory,  
Ann.Phys.} \textbf{235} (1994) 435 [hep-th/9312059].


\bibitem{Schaller} P. Schaller, T. Strobl,
\textit{Poisson structure induced (topological) field theories,
Modern Phys. Lett.} \textbf{A 9} (1994) 3129 [hep-th/9405110].

\bibitem{Str} P. Schaller, T. Strobl,
\textit{Introduction to Poisson $\sigma$-models},
In \textit{Low-Dimensional Models in Statistical Physics and Quantum Field Theory}, Lecture Notes in Physics \textbf{469}, Springer (1996) 321 [hep-th/9507020].



\bibitem{Birm} D. Birmingham, M. Blau, M. Rakowski, G. Thompson,
\textit{Topological Field Theories},
\textit{Phys. Rep.} \textbf{209} (1991) 129.

\bibitem{Cheeger-Simons} J. Cheeger, J. Simons,
\textit{Differential characters and geometric invariants},
in J. Alexander, J. Harer (eds.) \textit{Geometry and Topology, Proceedings of the Special Year held at the University of Maryland, College Park 1983–1984}, Lect. Notes Math. \textbf{1167}, Springer Berlin Heidelberg (1985).



\bibitem{CS} S. Chern, J. Simons,
\textit{Characteristic forms and geometric invariants, 
Annals of Mathematics} \textbf{99} (1974) 48.

\bibitem{WittenDij} R. Dijkgraaf, E. Witten,
\textit{Topological gauge theories and group cohomology,
Comm. Math. Phys.} \textbf{129} (1990) 393.

\bibitem{Milnor}  J.W. Milnor and J. D. Stasheff, 
\textit{Characteristic Classes}, Princeton University Press (1974).


\bibitem{Migdal}  A. Migdal, 
\textit{Recursion Relations in Gauge Theories, Zh. Eksp. Teor. Fiz.} \textbf{69} (1975) 810 (\textit{Sov. Phys. Jetp.} \textbf{42} 413).

\bibitem{Bralic} N. Bralic, 
\textit{Exact Computation of Loop Averages in Two-Dimensional Yang-Mills Theory, 
Phys. Rev.} \textbf{D 22} (1980) 3090.

\bibitem{Kazak} V. Kazakov, I. Kostov, 
\textit{Non-linear Strings in Two-Dimensional $U(\infty)$ Gauge Theory, 
Nucl. Phys.} \textbf{B 176} (1980) 199. 

\bibitem{Kazak2} V. Kazakov, I. Kostov, 
\textit{Computation of the Wilson Loop Functional in Two-Dimensional $U(\infty)$ Lattice Gauge Theory, 
Phys. Lett.} \textbf{B 105} (1981) 453. 

\bibitem{Kazak3} V. Kazakov, 
\textit{Wilson Loop Average for an Arbitrary Contour in Two Dimensional U(N) Gauge Theory, 
Nuc. Phys.} \textbf{B 179} (1981) 283.

\bibitem{Gross} L. Gross, C. King, A. Sengupta, 
\textit{Two-Dimensional Yang-Mills via Stochastic Differential Equations, 
Ann. of Phys.} \textbf{194} (1989) 65.

\bibitem{Rus}  B. Rusakov, 
\textit{Loop Averages And Partition Functions in U(N) Gauge Theory On Two-Dimensional Manifolds, 
Mod. Phys. Lett.} \textbf{A 5} (1990) 693.

\bibitem{Fine} D. Fine, 
\textit{Quantum Yang-Mills On The Two-Sphere, 
Commun. Math. Phys.} \textbf{134} (1990) 273.

\bibitem{Fine2} D. Fine,
\textit{Quantum Yang-Mills On A Riemann Surface, 
Commun. Math. Phys.} \textbf{140} (1991) 321.

\bibitem{Blau2} M. Blau, G. Thompson,
\textit{Quantum Yang-Mills Theory On Arbitrary Surfaces},
\textit{Int. J. Mod. Phys.} \textbf{A 7} (1992) 3781.

\bibitem{Witten5} E. Witten,
\textit{On Quantum gauge theories in two dimensions,
Commun. Math. Phys.} \textbf{141} (1991) 153.

\bibitem{Moore} S. Cordes, G. Moore, S. Ramgoolam,
\textit{Lectures on 2D Yang-Mills Theory, Equivariant Cohomology and Topological Field Theories, 
Nucl. Phys. Proc. Suppl.} \textbf{41} (1995) 184 [hep-th/9411210].

\bibitem{Witten2} E. Witten,
\textit{Two Dimensional Gauge Theories Revisited,
J.Geom.Phys.} \textbf{9} (1992) 303 [hep-th/9204083].

\bibitem{Steen} N. Steenrod,
\textit{The topology of Fiber Bundles}, Princeton Mathematical Series \textbf{14},
Princeton University Press (1951).


\bibitem{Schweigert}
J. Fuchs, C. Schweigert, A. Valentino, 
\textit{A geometric approach to boundaries and surface defects in Dijkgraaf-Witten theories, }
arXiv:1307.3632.

\bibitem{Hatcher} A. Hatcher,
\textit{Algebraic Topology},
Cambridge University Press (2002).





 
























































































\end{thebibliography}
\end{document}